\documentclass[preprint,showpacs,preprintnumbers,amsmath,amssymb]{revtex4}

\usepackage{graphicx}% Include figure files
\usepackage{dcolumn}% Align table columns on decimal point
\usepackage{bm}% bold math
\usepackage{amssymb}

\begin{document}
\title{Specific heat anomalies of small quantum systems \\
subjected to finite baths
%Quantum small systems subjected to finite baths: \\
%Specific heat anomalies
}% Force line breaks with \\

\author{Hideo Hasegawa}
\altaffiliation{hideohasegawa@goo.jp}
\affiliation{Department of Physics, Tokyo Gakugei University,  
Koganei, Tokyo 184-8501, Japan}%

\date{\today}% It is always \today, today,

\begin{abstract}
We have studied the specific heat of the $(N_S+N_B)$ model
for an $N_S$-body harmonic oscillator (HO) system which is strongly coupled
to an $N_B$-body HO bath without dissipation.
The system specific heat of $C_S(T)$ becomes $N_S k_B$ 
at $T \rightarrow \infty$ and vanishes at $T = 0$ 
in accordance with the third law of thermodynamics. 
The calculated $C_S(T)$ at low temperatures
is not proportional to $N_S$ and shows an anomalous temperature dependence,
strongly depending on $N_S$, $N_B$ and the system-bath coupling.
In particular at very low (but finite) temperatures, it may become {\it negative} 
for a strong system-bath coupling, which is in contrast 
with {\it non-negative} specific heat of an HO system with $N_S=1$
reported by G-L. Ingold, P. H\"{a}nggi and P. Talkner
[Phys. Rev. E {\bf 79}, 061105 (2005)].
Our calculation indicates an importance of taking account of finite $N_S$ in studying 
open quantum systems which may include an arbitrary number of particles in general.

\end{abstract}

%\keywords{Fisher information, nonextensive statistics,
%spatial correlation}%Use showkeys class option if keyword
        
%05.30.-d  Quantum statistical mechanics
%05.70.-a  thermodynamics
%65.40.Ba  Heat capacity
%65.80.+n  Thermal properties of small particles, nanocrystals, and nanotubes
%
%05.40.-a  Fluctuation phenomenon, random process, noise
%05.10.Gg  Stochastic model
%05.45.-a  Nonlinear dynamics and chaos                          
%89.70.Cf  entropy and other measures of information

\maketitle
\newpage
\section{Introduction}
In recent years, there has been considerable interest in physical properties
of small-scale systems, which are prepared by advanced new techniques \cite{Ritort08}.
Stimulated by this development, a study on small-scale systems 
which is one of important areas in classical and quantum statistics,
has been extensively made \cite{Weiss99}.
Theoretical studies on open systems have been made with the use of
the Caldeira-Leggett (CL) type models 
\cite{Ullersma66,Caldeira81,Caldeira83,Ford87,Ford88}, 
in which a single particle is assumed to
be subjected to a bath consisting of uncoupled harmonic oscillators (HOs).
The CL model was originally proposed for an infinite bath ($N_B \rightarrow \infty $).
Recent studies, however, have employed the CL model with finite $N_B$
for studies of properties of small system coupled to finite bath 
\cite{Smith08,Wei09,Rosa08}.
Refs. \cite{Smith08,Wei09} have studied thermalization of a particle (the system) 
coupled to a finite bath, showing that a complete thermalization of the
particle requires some conditions for relative ranges of
oscillating frequencies in the system and bath.
The energy exchange between particles in a rachet potential
(the system) and finite bath ($N_B=1-500$) has been investigated
\cite{Rosa08}.

In CL-type models having been proposed for open systems 
\cite{Ullersma66,Caldeira81,Caldeira83,Ford87}, there are two issues
when they are applied to realistic open systems:
(a) the number of particles in a system is taken to be unity ($N_S=1$)
and (b) a system-bath coupling is assumed to be weak 
although the overall damping can be strong.
As for the issue (a), a number of particles in a system
is required to be finite since a generic open system may contain any number of particles.
CL-type models with $N_S=2$ have been investigated in Refs. \cite{Chou08,Gelin09}.
In our previous paper \cite{Hasegawa11a}, we proposed the ($N_S+N_B$) model 
in which a finite $N_S$-body system ($N_S \geq 1$) is coupled to finite $N_B$-body bath.
It has been shown that calculated energy distributions of a system 
show intrigue properties as functions of
$N_S$, $N_B$ and a system-bath coupling \cite{Hasegawa11a}.

As for the issue (b), Refs. \cite{Hanggi06,Hanggi08} have pointed out
ambiguities in defining physical quantities such as energy and specific heat
when a system-bath coupling is {\it not} weak.
Two different routes toward the evaluating of a system energy 
have been proposed for a system-plus-bath \cite{Hanggi06,Hanggi08}.
%for the Hamiltonian given by $H=H_S+H_B+H_I$
%where $H_S$ and $H_B$ denote contributions form a system and bath, respectively,
%and $H_I$ stands for the system-bath interaction.
The first route is based on the system partition function as given by
a ratio between the total and bath partition functions [Eq. (\ref{eq:C1})],
which is traditionally identified as the partition function of an open system
\cite{Grabert88}. 
The second route is based on the expectation value of the system Hamiltonian
averaged over the total Hamiltonian.
In the limit of vanishing or weak interaction, the two definitions yield
the same results. It is, however, not the case in general when the interaction is 
not weak (finite).
The specific heat of the system of a single free particle
coupled to a bath described by the Drude model
has been studied with the use of the two evaluation methods
\cite{Hanggi06,Hanggi08,Ingold09}.
Specific heats obtained by the two routes are different not only at low temperatures
but also in the leading high-temperature correction terms \cite{Hanggi06,Hanggi08}.
In particular, the specific heat in the first route 
is negative at low temperature while that in the second route is not.
In contrast, similar calculations for a single HO system subjected to 
a single bath oscillator or Drude bath have reported that the first route
does not yield a negative specific heat \cite{Ingold09}.
The difference between negative specific heat in a free particle system and 
non-negative one in an HO system within the first route is attributed to the fact that
the degree of freedom in the former is smaller than that of the latter \cite{Ingold09}. 
The obtained negative specific heat is not unphysical because the system specific heat
calculated by the first route should be interpreted as a change in the
specific heat of the environment when a system degree of freedom is attached
\cite{Hanggi06,Hanggi08,Ingold09}.
We should note that studies of Refs. \cite{Hanggi06,Hanggi08,Ingold09} have been made 
for the CL model with $N_S=1$. It is not clear whether conclusions obtained 
in Refs. \cite{Hanggi06,Hanggi08,Ingold09} are valid in a general case, 
for example, for finite $N_S$ ($> 1$).

It is interesting to study thermodynamical properties of
the $(N_S+N_B)$ model proposed in Ref. \cite{Hasegawa11a}, which is worthwhile 
for us to get some insight to the issues (a) and (b) and
which is the purpose of the present paper.
In conventional studies, deterministic dynamics of
particles in a system-plus-bath is replaced by the
stochastic Langevin equation.
By using the alternative method in this study, 
we will evaluate eigenfrequencies of the $(N_S+N_B)$ model 
where the system is not dissipative for finite $N_B$ \cite{Poincare}.
The energy and specific heat have been calculated
with the use of the first route for evaluating energy mentioned above.

The paper is organized as follows.
In Sec. II, we briefly explain the $(N_S+N_B)$ model 
for coupled HO system subjected to uncoupled HO bath \cite{Hasegawa11a},
emphasizing a difficulty in solving the $N_S$-coupled quantum Langevin equations.
With the use of canonical transformation, we derive the feasible quantum
Langevin equation, with which the expression for eigenfrequencies of the model is obtained.
Eigenfrequencies are analytically evaluated for a bath described by 
the identical-frequency model ($\omega_n=\omega_0$ for $n=1$ to $N_B$) \cite{Hasegawa11b}.
Thermodynamical quantities are expressed in terms of obtained eigenfrequencies.
We have made numerical calculations of temperature dependences of
the energy and specific heat, changing parameters of $N_S$, $N_B$ and
a system-bath coupling. An origin of the negative specific heat
at low temperatures is illustrated.
In Sec. III we study similar two models for uncoupled HO system subjected to
uncoupled HO bath (Sec. IIIA) and for coupled HO system subjected 
to coupled HO bath (Sec. IIIB). The final Sec. IV is devoted to our conclusion.

%\newpage
\section{Adopted ($N_S+N_B$) model}

\subsection{Coupled quantum Langevin equations}
We consider the $(N_S+N_B$) model in which 
the a one-dimensional $N_S$-body system ($H_S$) is subjected to 
an $N_B$-body bath ($H_B$) by the interaction ($H_{I}$) \cite{Hasegawa11a}. 
The total Hamiltonian is assumed to be given by
\begin{eqnarray}
H &=& H_S+H_B+H_{I},
\label{eq:A0}
\end{eqnarray}
with
\begin{eqnarray}
H_S &=& \sum_{k=1}^{N_S} \left[ \frac{P_k^2}{2M}+ \frac{D Q_k^2}{2}
+ \frac{K}{2}(Q_{k}-Q_{k+1})^2 \right],
%- F(t) \sum_{k=1}^{N_S} Q_k, 
\label{eq:A2}\\
H_B &=& \sum_{n=1}^{N_B} \left( \frac{p_n^2}{2m}
+ \frac{m \omega_n^2 q_n^2}{2} \right),
\label{eq:A3}\\
H_{I} &=&  \sum_{k=1}^{N_S}  \sum_{n=1}^{N_B} \frac{c_{kn}}{2} (Q_k-q_n)^2,
\label{eq:A4}
%\hspace{1cm}\mbox{(the model A)},
\end{eqnarray}
where $P_k$ ($p_n$) and $Q_k$ ($q_n$) express the momentum and position
operators, respectively, of an HO with mass $M$ ($m$) in the system (bath), 
$D$ and $K$ denote force constants in the system, 
and $c_{kn}$ is a system-bath coupling.
The system is subjected to a bath given by Eq. (\ref{eq:A3}) consisting of a collection of
uncoupled HOs with oscillator frequencies of $\{ \omega_n \}$.
Operators satisfy commutation relations,
\begin{eqnarray}
\left[Q_k, P_{\ell} \right] &=& i \hbar \delta_{k \ell}, \;\;\;
[q_n, p_m] = i \hbar \delta_{n m}, \;\;\;
[Q_k, Q_{\ell}]=[P_k, P_{\ell}]=[q_n, q_m]=[p_n, p_m]=0.
\end{eqnarray}
In a general case of $D \neq 0$ and $K \neq 0$, Eq. (\ref{eq:A2}) expresses the system
including coupled HOs. In the limit of $D \neq 0$ and $K=0$,
the system consists of a collection of uncoupled (independent) HOs.

In conventional approaches to the system-plus-bath model, 
we derive the quantum Langevin equation, employing the Heisenberg equation,
\begin{eqnarray}
i \hbar \dot{O} &=& [O, H],
\end{eqnarray} 
where $O$ stands for an arbitrary operator and a dot denotes a derivative 
with respect to time.
Equations of motion for $Q_k$ and $q_n$ are given by
\begin{eqnarray}
M \ddot{Q}_k &=& -D Q_k -K(2 Q_k-Q_{k-1}-Q_{k+1})
- \sum_{n=1}^{N_B} c_{kn} (Q_k-q_n), 
\label{eq:A5}\\
m \ddot{q}_n &=& -m \tilde{\omega}_n^2 q_n +\sum_{k=1}^{N_S} c_{kn} Q_k,
\label{eq:A6}
\end{eqnarray}
with
\begin{eqnarray}
m \tilde{\omega}_n^2 &=& m \omega_n^2 + \sum_{k=1}^{N_S} c_{kn}.
\label{eq:A7}
\end{eqnarray}
Substituting a formal solution of $q_n(t)$ into Eq. (\ref{eq:A5}), 
we obtain the quantum Langevin equations given by \cite{Hasegawa11a}
\begin{eqnarray}
M \ddot{Q}_k(t) &=& -D Q_k(t) -K\left[ 2Q_k(t)-Q_{k-1}(t)-Q_{k+1}(t) \right]
- M \sum_{\ell=1}^{N_S} \xi_{k \ell} Q_{\ell}(t) \nonumber \\
&-& \sum_{\ell=1}^{N_S} \int_0^t \gamma_{k \ell}(t-t') \dot{Q}_{\ell}(t')\:dt'
-\sum_{\ell=1}^{N_S} \gamma_{k \ell} Q_{\ell}(0)+\zeta_k(t)
\hspace{0.5cm}\mbox{for $k=1$ to $N_S$},
\label{eq:A8}
\end{eqnarray}
with
\begin{eqnarray}
M \xi_{k\ell} &=& \sum_{n=1}^{N_B}  \left( c_{kn}  \delta_{k \ell} 
-\frac{c_{kn} c_{\ell n}}{m \tilde{\omega}_n^2} \right), 
\label{eq:A9}\\
\gamma_{k\ell}(t) 
&=&\sum_{n=1}^{N_B} \left( \frac{c_{kn} c_{\ell n}}{m \tilde{\omega}_n^2}\right) 
\cos \tilde{\omega}_n t, 
\label{eq:A10}\\
\zeta_k(t) &=& \sum_{n=1}^{N_B} c_{kn} 
\left(q_n(0) \cos \tilde{\omega}_n t
+ \frac{ \dot{q}_n(0)}{\tilde{\omega}_n} \sin \tilde{\omega}_n t \right).
\label{eq:A11}
\end{eqnarray}
Here $\xi_{k \ell}$ denotes the additional interaction between
$k$ and $\ell$th particles in the system induced by couplings $\{ c_{kn} \}$, 
$\gamma_{k\ell}(t)$ stands for the memory kernel and $\zeta_k$ is the stochastic force.
By using averages in initial values of $q_n(0)$ and $\dot{q}_n(0)$,
\begin{eqnarray}
\langle m \tilde{\omega}_n^2 q_n(0)^2\rangle_B
&=& m \langle \dot{q}_n(0)^2 \rangle_B 
=\left( \frac{\hbar \tilde{\omega}_n }{2} \right)
\coth\left( \frac{\beta \hbar \tilde{\omega}_n}{2} \right),
%k_B T,
\label{eq:A12}
\end{eqnarray} 
we obtain the fluctuation-dissipation relation,
\begin{eqnarray}
\frac{1}{2}\langle \zeta_k(t)\zeta_{\ell}(t')  +\zeta_{\ell}(t') \zeta_k(t) \rangle_B &=& 
\sum_{n=1}^{N_B} \left( \frac{c_{kn} c_{\ell n}}{m \tilde{\omega}_n^2} \right)
\left( \frac{\hbar \tilde{\omega}_n }{2} \right)
\coth\left( \frac{\beta \hbar \tilde{\omega}_n}{2} \right)
\cos \tilde{\omega}_n (t-t'), \\
&\rightarrow& k_B T \gamma_{k \ell}(t-t')
\hspace{1cm}\mbox{for $\beta \rightarrow 0$},
\label{eq:A13}
\end{eqnarray}
where $\langle \cdot \rangle_B$ signifies the average over initial states of the bath. 
In the case of $N_S=1$, $\xi_{k k}$ in Eq. (\ref{eq:A9}) expressing a shift of oscillator
frequency due to an introduced coupling, vanishes 
if we adopt $c_{n}=m \tilde{\omega}_n^2$ \cite{Caldeira81,Caldeira83}.
In the case of $N_S \neq 1$, however, it is impossible to choose
$\{c_{kn} \}$ such as $\xi_{k \ell}=0$ for all pairs of $(k, \ell)$,
then $Q_k$ is inevitably coupled with $Q_{\ell}$ ($\ell \neq k$).
Because of these couplings between HOs, the $N_S$-body system
cannot be simply regarded as a sum of systems with $N_S=1$.
Although the quantum Langevin equations given by Eqs. (\ref{eq:A8})-(\ref{eq:A11}) 
are formally exact, it is difficult to solve them because they 
are given by $N_S$-coupled integrodifferential equations.

\subsection{Quantum Langevin equation with canonical transformation}
It is possible to derive the quantum Langevin equation which has a simpler
structure than that given by Eq. (\ref{eq:A8})-(\ref{eq:A11}).
We assume that $N_S$ is even without a loss of generality. 
Imposing a periodic boundary condition,
\begin{eqnarray}
Q_{N_S+k} &=& Q_{k}, \;\;\; P_{N_S+k}=P_{k},
\label{eq:B1}
\end{eqnarray}
we employ the canonical transformation \cite{Florencio85},
\begin{eqnarray}
Q_k &=& \frac{1}{\sqrt{N_S}} \sum_{s=-N_S/2}^{N_S/2-1}
e^{i (2 \pi k s/N_S)} \tilde{Q}_s, 
\label{eq:B2}\\
P_k &=& \frac{1}{\sqrt{N_S}} \sum_{s=-N_S/2}^{N_S/2-1}
e^{i (2 \pi k s/N_S)} \tilde{P}_s.
\label{eq:B3} 
\end{eqnarray}
Note that the boundary condition is satisfied in Eqs. (\ref{eq:B2}) and (\ref{eq:B3}) and 
that the set $\{ (1/\sqrt{N_S}) \:e^{i(2 \pi k/N_S) s} \}$ is orthogonal and complete
in a periodic domain of the oscillator label $k$ \cite{Florencio85}.
With the canonical transformation given by Eqs.(\ref{eq:B2}) and (\ref{eq:B3}), 
$H_S$ in Eq. (\ref{eq:A2}) becomes
\begin{eqnarray}
H_S &=& \sum_{s=-N_S/2}^{N_s/2-1} \left[\frac{\tilde{P}_s^* \tilde{P}_s}{2M}
+ \frac{M \Omega_s^2 \: \tilde{Q}_s^* \tilde{Q}_s}{2} \right], 
\label{eq:B4}
\end{eqnarray}
with
\begin{eqnarray}
M \Omega_s^2 &=& D+ 4K \sin^2 \left(\frac{\pi s}{N_S} \right)
\hspace{1cm}
\mbox{for $s=-\frac{N_S}{2}, -\frac{N_S}{2}+1, \cdot\cdot, \frac{N_S}{2}-1$},
\label{eq:B6}
\end{eqnarray}
where the commutation relations: 
\begin{eqnarray}
[\tilde{Q}_s, \tilde{P}_{s'}^*] = i \hbar \delta_{s s'},
\;\;\;[\tilde{Q}_s, \tilde{Q}_{s'}^*]=[\tilde{P}_s, \tilde{P}_{s'}^*]=0,
\end{eqnarray}
hold with $\tilde{Q}_s^*=\tilde{Q}_{-s}$ and $\tilde{P}_s^*=\tilde{P}_{-s}$.

The canonical transformation given by Eqs.(\ref{eq:B2}) and (\ref{eq:B3}) 
with an assumption,
\begin{eqnarray}
c_{kn}=c_n,
\end{eqnarray}
leads to $H_I$ given by
\begin{eqnarray}
H_I &=& \sum_{n=1}^{N_B} \left( \frac{c_n}{2} \right) 
\sum_{s=-N_S/2}^{N_S/2-1} \tilde{Q}_s^*\tilde{Q}_s
+\frac{N_S}{2} \sum_{n=1}^{N_B} c_n q_n^2
-\sqrt{N_S}\tilde{Q}_0 \sum_{n=1}^{N_B} c_n q_n.
\label{eq:B5}
\end{eqnarray}

From Eqs. (\ref{eq:A3}), (\ref{eq:B4}) and (\ref{eq:B5}), 
we may derive equations of motion for $\tilde{Q}_s$ and $q_n$,
\begin{eqnarray}
M \ddot{\tilde{Q}}_s &=& - M \tilde{\Omega}_s^2 \tilde{Q}_s 
\hspace{4cm} \mbox{for $s \neq 0$}, 
\label{eq:B7}\\
M \ddot{\tilde{Q}}_0 &=& - M \tilde{\Omega}_0^2 \tilde{Q}_0
+\sqrt{N_S} \sum_{n=1}^{N_B} c_n  q_n
\hspace{1cm} \mbox{for $s=0$}, 
\label{eq:B8}\\
m \ddot{q}_n &=& - m \tilde{\omega}_n^2 q_n + \sqrt{N_S} c_n \tilde{Q}_0,
\label{eq:B9}
\end{eqnarray}
with
\begin{eqnarray}
M \tilde{\Omega}_s^2 &=& M \Omega_s^2
%+ 4 K \sin^2 \left(\frac{\pi s}{N_S} \right)
+ \sum_{n=1}^{N_B} c_n
\hspace{0.5cm}
\mbox{for $s=-\frac{N_S}{2}, -\frac{N_S}{2}+1, \cdot\cdot, \frac{N_S}{2}-1$}, 
\label{eq:B10}\\
m \tilde{\omega}_n^2 &=& m \omega_n^2 + N_S c_n,
\label{eq:B11}
\end{eqnarray}
which show that effective frequencies of $\tilde{\Omega}_s$ and $\tilde{\omega}_r$ 
are increased with increasing the system-bath coupling $c_n$.

Substituting a formal solution of $q_n(t)$ of Eq. (\ref{eq:B9})
into Eq. (\ref{eq:B8}), we obtain
\begin{eqnarray}
M \ddot{\tilde{Q}}_0(t) 
&=& -\left[M \tilde{\Omega}_0^2 -\gamma(0) \right] \tilde{Q}_0(t)
- \int_{0}^{t} \gamma(t-t') \dot{\tilde{Q}}_0(t')\:dt'+ \zeta(t) 
-\gamma(t) \tilde{Q}_0(0),
\label{eq:G2}
\end{eqnarray}
where
\begin{eqnarray}
%\mu(t-t') &=& N_S \sum_{n} \left( \frac{ c_n^2}{m \tilde{\omega}_n} \right)
%\sin \tilde{\omega}_n (t-t') = \frac{d}{d t'} \gamma(t-t'), \\
\gamma(t-t') &=& N_S \sum_{n=1}^{N_B} \left( \frac{ c_n^2}{m \tilde{\omega}_n^2} \right)
\cos \tilde{\omega}_n (t-t'), 
\label{eq:G4}\\
\zeta(t) &=& \sqrt{N_S} \sum_{n=1}^{N_B} c_n
\left[ q_n(0) \cos \tilde{\omega}_n t 
+ \frac{\dot{q}_n(0)}{\tilde{\omega}_n} \sin \tilde{\omega}_n t \right].
\label{eq:G5}
\end{eqnarray}
Equations (\ref{eq:G2})-(\ref{eq:G5}) express 
the desired quantum Langevin equation for $\tilde{Q}_0$,
an $s = 0$ component of $\tilde{Q}_s$,
which has simpler structure than those given by Eq. (\ref{eq:A8})-(\ref{eq:A11}).
Note that solutions of $\tilde{Q}_s$ with $s \neq 0$ are obtained from Eq. (\ref{eq:B7}).
Once $\tilde{Q}_s$ for all $s$ are obtained, $Q_k$ is obtainable with the use of
the canonical transformation given by Eq. (\ref{eq:B2}).

We note that the quantum Langevin equation given by Eq. (\ref{eq:G2})
is similar to that in the conventional CL-type model 
with $N_S=1$ where $q_n$ couples to a single particle in the system.
We may employ various methods such as the quantum Langevin and master equations,
which have been widely adopted for the CL-type model.
In the case of $N_B \rightarrow \infty$, a sum in
the kernel given by Eq. (\ref{eq:G4}) is converted into the integral and the kernel 
$\gamma(t)$ may be approximated by the Drude or Ohmic one.

\subsection{Eigenfrequencies of small open systems}
In the present study, we pay our attention to small systems subjected to
non-dissipative finite bath \cite{Poincare}.
Equation (\ref{eq:G2})-(\ref{eq:G5}) may be rewritten as
\begin{eqnarray}
M \ddot{\tilde{Q}}_0(t) &=& -M \tilde{\Omega}_0^2 \tilde{Q}_0(t) 
+ \int_{0}^{t} \mu(t-t') \tilde{Q}_0(t') \:dt' + \zeta(t), 
\label{eq:G1}
\end{eqnarray}
with
\begin{eqnarray}
\mu(t-t') &=& N_S \sum_{n=1}^{N_B} \left( \frac{ c_n^2}{m \tilde{\omega}_n} \right)
\sin \tilde{\omega}_n (t-t') = \frac{d}{d t'} \gamma(t-t'),
\label{eq:G3}
\end{eqnarray}
which are more tractable than Eqs. (\ref{eq:G2})-(\ref{eq:G5}) 
for finite non-dissipative baths.
Applying the Laplace transformation to Eqs. (\ref{eq:G1}) and (\ref{eq:G3}),
\begin{eqnarray}
\hat{Q}_0(z) &=& \int_0^{\infty} e^{-zt} \tilde{Q_0}(t) \:dt, 
\label{eq:B12}\\
\hat{q}_n(z) &=& \int_0^{\infty} e^{-zt} q_n(t) \:dt, 
\label{eq:B13}
\end{eqnarray}
we obtain
\begin{eqnarray}
\hat{Q}_0(z)&=&  \hat{G}(z)  \left[\dot{\tilde{Q}}_0(0)+ z \tilde{Q}_0(0) 
+  \sum_{n=1}^{N_B} 
 \frac{\sqrt{N_S} \:c_n}{M}\left( \frac{\dot{q}_n(0)+ z q_n(0)}{z^2+\tilde{\omega}_n^2} \right)
\right], 
\label{eq:B14}  
\end{eqnarray}
with
\begin{eqnarray}
\hat{G}(z) &=& \left[z^2+\tilde{\Omega}_0^2 
- N_S \sum_{n=1}^{N_B} \frac{c_n^2 }{m M(z^2+\tilde{\omega}_n^2)}\right]^{-1}.
\label{eq:B15}
\end{eqnarray}

We note in Eqs. (\ref{eq:B7})-(\ref{eq:B9}) that $\tilde{Q}_s$ for $s \neq 0$ are 
decoupled from the rest of variables, while $\tilde{Q}_0$ is coupled with $q_n$. 
($N_S-1$) eigenfrequencies of $\tilde{Q}_s(t)$ are given by $\tilde{\Omega}_s$ 
in Eq. (\ref{eq:B7}). Remaining $(N_B+1)$ eigenfrequencies are given by roots
of $\hat{G}(-i \nu)^{-1}=0$,
\begin{eqnarray}
\nu^2-\tilde{\Omega}_0^2 - N_S \sum_{n=1}^{N_B} \frac{c_n^2}{m M(\nu^2-\tilde{\omega}_n^2)}
&=& 0.
\label{eq:B16}
\end{eqnarray}

Alternatively, we may obtain  $(N_B+1)$ eigenfrequencies as follows:
Calculating the determinant derived from Eqs. (\ref{eq:B8}) and (\ref{eq:B9})
which is expressed in a basis of $(s=0, n=1, 2,3, \ldots)$,
\[ \left|
 \begin{array}{ccccc}
  M(\tilde{\Omega}_0^2 - \nu^2) &  -c_1 \sqrt{N_S} &  -c_2 \sqrt{N_S}
  & -c_3 \sqrt{N_S}& \ldots \\
  -c_1 \sqrt{N_S}   & m(\tilde{\omega}_1^2-\nu^2) & 0& 0 & \ldots \\ 
  -c_2 \sqrt{N_S}   &0  & m(\tilde{\omega}_2^2-\nu^2) & 0 &\ldots \\
  -c_3 \sqrt{N_S}   &0  &0 & m(\tilde{\omega}_3^2-\nu^2) & \ldots \\ 
  \ldots &  \ldots & \ldots & \ldots & \ldots     
 \end{array} 
\right|=0, \]
and using a sweeping method to make a triangle determinant, 
we obtain an equation for eigenfrequencies,
\begin{eqnarray}
\left[M(\tilde{\Omega}_0^2-\nu^2) - 
\sum_{n=1}^{N_B} \frac{c_n^2 N_S}{m(\tilde{\omega}_n^2-\nu^2)} \right]
\prod_{n=1}^{N_B} m(\tilde{\omega}_n^2-\nu^2) =0, 
\label{eq:B17}
\end{eqnarray}
which is equivalent to Eq. (\ref{eq:B16}).

It is difficult to analytically solve Eq. (\ref{eq:B16}) or (\ref{eq:B17}) 
in a general case of $\{\omega_n \}$, which requires numerical methods.
However, when we adopt the identical-frequency model for the bath \cite{Hasegawa11b},
\begin{eqnarray}
\omega_n &=& \omega_0,\;\;\;c_n = c 
\hspace{1cm}\mbox{for $n=1$ to $N_B$},
\label{eq:B19}
\end{eqnarray}
we may easily obtain eigenfrequencies of $\nu_i$ ($i=1$ to $N_S+N_B$),
\begin{center}
\begin{tabular}{|c|| c|c|c|c|c|c|c|c|c|c|} \hline
$\;\;i\;\;$ & $1$ &$\cdots$ & $N_S/2+1$ & $\cdots$ & $N_S$ & $N_S+1$ 
& $\cdots$ & $N_S+N_B/2+1$ & $\cdots$ & $N_S+N_B$ \\ \hline
$\;\;\nu_i^2\;\;$ & $\tilde{\Omega}_{-N_S/2}^2$ & $\cdots$ & $\nu_{+}^2$ & $\cdots$
& $\tilde{\Omega}_{N_S/2-1}^2$ & $\tilde{\omega}_0^2$ & $\cdots$ 
& $\nu_{-}^2$ & $\cdots$ & $\tilde{\omega}_0^2$ \\ \hline
\end{tabular}
\end{center}
with
\begin{eqnarray}
\nu_{\pm}^2 &=&
\frac{1}{2}\left[ \tilde{\Omega}_0^2+\tilde{\omega}_0^2
\pm \sqrt{(\tilde{\Omega}_0^2-\tilde{\omega}_0^2)^2+ \frac{4 N_S N_B c^2}{M m}}
\right],
\label{eq:B18}\\
M \tilde{\Omega}_s^2 &=& M \Omega_s^2+N_B \:c, \\
m \tilde{\omega}_0^2 &=& m \omega_0^2+ N_S \:c, 
\end{eqnarray}
where $\Omega_s$ is given by Eq. (\ref{eq:B6}).
It is easy to see that in the limit of no couplings ($c=0$), eigenfrequencies 
are given by
\begin{center}
\begin{tabular}{|c|| c|c|c|c|c|c|c|c|c|c|} \hline
$\;\;i\;\;$ & $1$ &$\cdots$ & $N_S/2+1$ & $\cdots$ & $N_S$ & $N_S+1$ 
& $\cdots$ & $N_S+N_B/2+1$ & $\cdots$ & $N_S+N_B$ \\ \hline
$\;\;\nu_i^2\;\;$ & $\Omega_{-N_S/2}^2$ & $\cdots$ & $\Omega_0^2$ & $\cdots$
& $\Omega_{N_S/2-1}^2$ & $\omega_0^2$ & $\cdots$ 
& $\omega_0^2$ & $\cdots$ & $\omega_0^2$ \\ \hline
\end{tabular}
\end{center}

In the case of $N_S=N_B=1$ with $K=0$, $D=f_S$, $\omega_0=0$ and $c=f_B$,
Eq. (\ref{eq:B18}) yields eigenfrequencies which are equivalent with those
expressed by Eq. (17) of Ref. \cite{Ingold09},
related discussion being given in Appendix A.
In a pedagogical case of $N_S=2$ and $N_B=1$, we may exactly solve a system-plus-bath,
transforming original variables of $Q_k$ and $P_k$ $(k=1,2)$
into center-of-mass and relative variables.
Obtained eigenfrequencies agree with those given by Eq. (\ref{eq:B18}) 
as is shown in Appendix A.

\subsection{System partition function, energy and specific heat}

It has been shown in the preceding subsection that the system-plus-bath may be effectively 
expressed as a collection of independent HOs with eigenfrequencies $\{ \nu_i \}$.
The system partition function is expressed in terms of eigenfrequencies by
\begin{eqnarray}
Z_S &=& \frac{Z}{Z_B},
\label{eq:C1}
\end{eqnarray}
with
\begin{eqnarray}
Z &=& {\rm Tr} \;e^{-\beta H}
= \prod_{i=1}^{N_S+N_B} \left[ \frac{1}{2 \sinh(\beta \hbar \nu_i/2 )} \right], 
\label{eq:C2}\\
Z_B &=& {\rm Tr}_B \;e^{-\beta H_B}
= \prod_{j=1}^{N_B} \left[\frac{1}{2 \sinh (\beta \hbar \omega_j/2)} \right],
\label{eq:C3b}
\end{eqnarray}
where Tr and ${\rm Tr}_B$ denote a full trace over all variables and
a partial trace over bath variables, respectively, and 
a bath frequency is given by $\omega_j=\omega_0$.
The energy and specific heat of the system are given by
\begin{eqnarray}
E_S &=& -\frac{\partial \ln Z_S}{\partial \beta}, \\
&=& \sum_{i=1}^{N_S+N_B} \left(\frac{\hbar \nu_i}{2} \right) 
{\rm coth}\left(\frac{\beta \hbar \nu_i}{2} \right)
-  \left(\frac{N_B \hbar \omega_0}{2} \right) 
{\rm coth}\left(\frac{\beta \hbar \omega_0}{2} \right), 
\label{eq:C3}\\
C_S &=& k_B \sum_{i=1}^{N_S+N_B} 
\left[\frac{\beta \hbar \nu_i}
{2 \sinh (\beta \hbar \nu_i/2 )} \right]^2
- k_B N_B \left[\frac{\beta \hbar \omega_0}
{2 \sinh (\beta \hbar \omega_0/2 )} \right]^2
\label{eq:C4},\\
&\equiv & C-C_B,  
\label{eq:C4b}
\end{eqnarray}
where the first term of $C$ expresses the specific heat of a system-plus-bath 
and the second term of $C_B$ denotes the specific heat of the bath alone.  
In the zero- and high-temperature limits, $E_S$ and $C_S$ become
\begin{eqnarray}
E_S &=& \left\{ \begin{array}{ll}
\sum_{i=1}^{N_S+N_B} 
\left(\frac{\hbar \nu_i}{2} \right) 
- \sum_{j=1}^{N_B} \left(\frac{\hbar \omega_j}{2} \right)
\quad & \mbox{for $k_B T=0$}, 
\label{eq:C5}\\ 
N_S k_B T
\quad & \mbox{for $k_B T \gg \hbar \nu_i, \hbar w_0$},
\end{array} \right. 
\label{eq:C6}\\ 
%\end{eqnarray}
%\begin{eqnarray}
C_S &=& \left\{ \begin{array}{ll}
0
\quad & \mbox{for $k_B T=0$}, \\ 
N_S k_B
\quad & \mbox{for $k_B T \gg \hbar \nu_i, \hbar w_0$}.
\end{array} \right. 
\label{eq:C7}
\end{eqnarray}

\subsection{Model calculations for coupled HO systems 
subjected to uncoupled HO baths}

We have made numerical calculations of $E_S$ and $C_S$,
by using Eqs. (\ref{eq:C3}) and (\ref{eq:C4}).
The system-bath coupling $c$ is assumed to be given by \cite{Hasegawa11a}
\begin{eqnarray}
c &=& \frac{c_0}{N_S N_B},
\end{eqnarray}
such that the interaction term in Eq. (\ref{eq:A4}) including summations over
$\sum_{k=1}^{N_S}$ and $\sum_{n=1}^{N_B}$ yield finite 
contributions even in the limits of $N_S \rightarrow \infty$
and/or $N_B \rightarrow \infty$. Our $(N_S+N_B)$ model includes 
eight parameters: $M$, $D$, $K$, $N_S$, $m$, $\omega_0$, $N_B$ and $c_0$.
Bearing in mind coupled HO system subjected to uncoupled HO bath,
we have employed $K=1.0$, $D=0.0$, $m=1.0$, $M=1.0$ and $\omega_0=1.0$ 
otherwise noticed, related discussion being given in Sec. III.
A unit of energy of the model is given by $\epsilon_0=\hbar \omega_0$.

\subsubsection{Isolated system and bath}
First we show calculated results of isolated system and bath 
with $c_0=0.0$.
Figure \ref{fig1}(a) shows $E_B/N_B$ and $E_S/N_S$ for $N_S=4$, 10, 100 and 1000.
The bath energy is $E_B/N_B=\hbar \omega/2$ at $T=0$ and gradually approaches
$k_B T$ at high temperatures. The $N_S$ dependence of the system energy $E_S/N_S$ is
not so evident except for very low temperature (below).
Figure \ref{fig1}(b) shows $C_B/k_B N_B$ and $C_S/k_B N_S$ for various $N_S$.
The bath specific heat $C_B/k_B N_B$ follows Einstein's formula showing exponential
decrease at low temperatures.
In contrast, the system specific heat becomes
$C_S/k_B N_S=1/N_S$ at $T \simeq 0.0$, although it reduces to zero at $T=0$
as shown by Eq. (\ref{eq:C7}a). This transition is more clearly seen 
when we introduce an infinitesimal coupling (Fig. \ref{fig2}).
%At $0.25 < k_B T < 0.5$, 
At high temperatures, $C_S/k_B N_S$ approaches unity as expected.
In the intermediate temperature range, $C_S$ nearly follows the linear $T$.
The linear-$T$ specific heat is examined in the inset of Fig. \ref{fig1}(b)
where $C_S/k_B T N_S$ is plotted as a function of $T$.
We note that $C_S/k_B T N_S$ for $N_S=100$ and $1000$ is almost constant
at $0.01 < T < 0.6$.
It is noted that the linear-$T$ behavior of the calculated specific heat 
is a consequence of the one-dimensional coupled HO model adopted in this study. 
If we employ the three-dimensional coupled HO model for the system, 
we obtain the $T^3$-specific heat at low temperatures. % as the Debye model shows.

\subsubsection {Effects of $c_0$}
Figure \ref{fig2} shows $C_S/k_B N_S$ for various coupling strengths
with $N_S=4$ and $N_B=100$.
The system specific heat for $c_0=0.0$ is $C_S/k_B N_S = 1/N_S$
at $T \simeq 0.0$ as mentioned above.
When a small interaction is introduced, $C_S$ clearly reduces to 
zero at $T=0.0$. With increasing the coupling strength, $C_S$ 
is furthermore decreased at low temperatures where it shows
an anomalous temperature dependence.
For $c_0 \geq 2.0$, the specific heat becomes negative, and 
magnitudes of negative dips are increased with increasing $c_0$.

\subsubsection {Effects of $N_S$}
The $N_S$ dependent specific heat is shown in Figs. \ref{fig3}(a), (b) and (c) where
$C_S/k_B N_S$ is plotted for $c_0=0.1$, 1.0 and 5.0, respectively, for various $N_S$
with a fixed $N_B=100$.
%for $N_S=4$ (solid curve), 6 (dotted curve), 10 (dashed curve) 
%and 100 (chain curve) .
Figure \ref{fig3}(a) shows that for $c_0=0.1$ an anomalous bump in $C_S$ at low temperatures
is gradually decreased with increasing $N_S$ and it well follows the linear-$T$ behavior
at higher temperatures.
For $c_0=1.0$ the temperature dependence of the specific heat
is almost independent of $N_S$ as shown by Fig. \ref{fig3}(b).
In contrast, Fig. \ref{fig3}(c) shows that
$C_S$ strongly depends on $N_S$ for $c_0=5.0$, for which magnitude of
negative specific heat is much increased for smaller $N_S$.

\subsubsection {Effects of $N_B$} 
The $N_B$ dependence of the system specific heat is shown 
in Fig. \ref{fig4}, where $C_S/k_B N_S$ is plotted
for $c_0=0.1$, 1.0 and 5.0 with various $N_B$ and a fixed $N_S=4$.
%for $N_B=4$ (solid curve), 10 (dashed curve), 100 (dotted curve) and 1000 (chain curve)
%with $c_0=0.1$ and $c_0=1.0$ .
In the case of $c_0=0.1$, the calculated specific heats are almost independent of $N_B$.
In the case of $c_0=1.0$, a bump in $C_S$ is gradually decreased
with increasing $N_B$. 
On the contrary in the case of $c_0=5.0$, magnitudes of negative dips in
the specific heat are more significant with increasing $N_B$ 
although the result for $N_B=1000$ is nearly the same as that for $N_B=100$. 

\subsubsection{Origin of the negative system specific heat}
We will elucidate the physical origin of the negative system specific heat
for a typical case of $N_S=4$, $N_B=10$ and $c_0=5.0$, whose
result has been presented in Fig. \ref{fig4}.
Chain and dashed curves in Fig. \ref{fig5} express
a total specific heat $C$ and a bath contribution $C_B$, respectively, which
arise from the first and second terms in Eq. (\ref{eq:C4}).
The solid curve denotes the system specific heat of $C_S$ $(=C-C_B)$,
which is given by the difference between the chain and dashed curves.
The inset of Fig. \ref{fig5} shows eigenfrequencies $\{ \omega_j \}$ 
for $c_0=0.0$ (open circles) and $c_0=5.0$ (filled circles).
In Eq. (\ref{eq:C4}) $C$ is expressed in terms of $\{ \omega_j \}$ 
for $c_0=5.0$ which become larger than those for $c_0=0.0$
by an introduced system-bath interaction.  
As a consequence, the specific heat of system-plus-bath is suppressed
at low temperatures compared to $C_B$ for the bath HOs
which is expressed in terms of $\omega_0$.
Then the system specific heat $C_S$ given by the difference of $C-C_B$ 
becomes negative at low temperatures. 
The obtained negative specific heat is not related with an instability of the system.

%\newpage
\section{Discussion}
\subsection{Uncoupled HO systems subjected to uncoupled HO baths}

%\subsubsection {Effects of $D$}
In the preceding section, we considered coupled HO system with $D=0.0$ and $K=1.0$. 
On the other hand, when we adopt $D=1.0$ and $K=0.0$, 
the model given by Eqs. (\ref{eq:A0})-(\ref{eq:A4}) expresses
the {\it uncoupled} HO system subjected to uncoupled HO bath.
Eigenfrequencies of $\nu_i$ ($i=1$ to $N_S+N_B$) are given by
\begin{center}
\begin{tabular}{|c|| c|c|c|c|c|c|c|c|c|c|} \hline
$\;\;i\;\;$ & $1$ &$\cdots$ & $N_S/2+1$ & $\cdots$ & $N_S$ & $N_S+1$ 
& $\cdots$ & $N_S+N_B/2+1$ & $\cdots$ & $N_S+N_B$ \\ \hline
$\;\;\nu_i^2\;\;$ & $\tilde{\Omega}_0^2$ & $\cdots$ & $\nu_{+}^2$ & $\cdots$
& $\tilde{\Omega}_0^2$ & $\tilde{\omega}_0^2$ & $\cdots$ 
& $\nu_{-}^2$ & $\cdots$ & $\tilde{\omega}_0^2$ \\ \hline
\end{tabular}
\end{center}
with
\begin{eqnarray}
\nu_{\pm}^2 &=&
\frac{1}{2}\left[ \tilde{\Omega}_0^2+\tilde{\omega}_0^2
\pm \sqrt{(\tilde{\Omega}_0^2-\tilde{\omega}_0^2)^2+ \frac{4 N_S N_B c^2}{M m}}
\right], \\
%\label{eq:B18}
M \tilde{\Omega}_0^2 &=& D + N_B \:c, \\
m \tilde{\omega}_0^2 &=& m \omega_0^2+ N_S \:c. 
\end{eqnarray}
In the limit of no couplings ($c=0$), eigenfrequencies 
are given by
\begin{center}
\begin{tabular}{|c|| c|c|c|c|c|c|c|c|c|c|} \hline
$\;\;i\;\;$ & $1$ &$\cdots$ & $N_S/2+1$ & $\cdots$ & $N_S$ & $N_S+1$ 
& $\cdots$ & $N_S+N_B/2+1$ & $\cdots$ & $N_S+N_B$ \\ \hline
$\;\;\nu_i^2\;\;$ & $\Omega_{0}^2$ & $\cdots$ & $\Omega_0^2$ & $\cdots$
& $\Omega_{0}^2$ & $\omega_0^2$ & $\cdots$ 
& $\omega_0^2$ & $\cdots$ & $\omega_0^2$ \\ \hline
\end{tabular}
\end{center}

The system specific heat calculated with the use of Eq. (\ref{eq:C4}) 
is shown in Fig. \ref{fig6}(a)
where $C_S/k_B N_S$ is plotted for various $N_S$ with $N_B=100$ and $c_0=5.0$.
For $c_0=0.0$, $C_S/k_B N_S$ follows the Einstein specific heat 
[see the dashed curve in Fig. 1(b)].
When $c_0$ is much increased, the specific heat becomes negative at low temperatures.
It is noted that the solid curve for $N_S=1$ and $N_B=100$ shows 
a negative $C_S$ at $0< k_B T/\epsilon_0 < 0.805$ ($\epsilon_0=\hbar \omega_0$),
while $C_S$ for $N_S=N_B=1$ is non-negative \cite{Ingold09}.
Figure 6(a) clearly shows
\begin{eqnarray}
C_S(T; N_S,N_B) &\neq& N_S\:C_S(T; 1,N_B)
\hspace{1cm}\mbox{for $T \simeq 0.0$},
\label{eq:X3} \\
&=&  N_S\:C_S(T; 1, N_B)
\hspace{1cm}\mbox{for $T \rightarrow \infty$},
\end{eqnarray}
where $C_S(T; N_S, N_B)$ denotes the system specific heat of 
the $(N_S+N_B)$ model at temperature $T$.

In order to examine an origin of the negative specific heat, we plot
$C$, $C_B$ and $C_S$ for $N_S=4$, $N_B=10$ and $c_0=5.0$ in Fig. \ref{fig6}(b),
whose inset shows eigenfrequencies for $c_0=0.0$ (open circles) and $c_0=5.0$ (filled circles).
We obtain a negative $C_S$ at $0 < k_B T/\epsilon_0 < 0.377$ where $C < C_B$
although $C_S > 0$ at $k_B T/\epsilon_0 > 0.377$.
Eigenfrequencies for $c_0=0.0$ are $\nu_i=1.0$ for $i=1$ to 14, and they are increased
by an introduced system-bath coupling of $c_0=5.0$ as shown in the inset.
From a comparison between Figs. 5 and 6, we note that a negative 
specific heat is realized both in coupled and uncoupled HO systems subjected 
to uncoupled HO baths, independently of eigenfrequencies of the system.

\subsection{Coupled HO systems subjected to coupled HO baths}

We have so far considered that a system is subjected
to a bath including a collection of {\it uncoupled} HOs. 
Here we will study a case in which a bath consists of {\it coupled} HOs.
The system-plus-bath is described by the Hamiltonian given 
by Eqs. (\ref{eq:A0})-(\ref{eq:A4}) but $H_B$ is replaced by
\begin{eqnarray}
H_B &=& \sum_{n=1}^{N_B} \left[ \frac{p_n^2}{2m}
+ \frac{k_n}{2}(q_n-q_{n+1})^2 \right],  
\label{eq:D1}
%\hspace{1cm}\mbox{(model B)},
\end{eqnarray}
where $k_n$ stands for a force constant between neighboring particles in the bath. 
We assume that $N_B$ is even and $k_n=k$ for $n=1$ to $N_B$,
imposing the periodic boundary condition given by
\begin{eqnarray}
q_{N_B+n} &=& q_{n}, \;\;\; p_{N_B+n}=p_{n}.
\label{eq:D2}
\end{eqnarray}
By using the canonical transformation with the identical-frequency model 
[Eq. (\ref{eq:B19})], we obtain eigenfrequencies given by (detail being given in Appendix B)
\begin{center}
\begin{tabular}{|c|| c|c|c|c|c|c|c|c|c|c|} \hline
$\;\;i\;\;$ & $1$ &$\cdots$ & $N_S/2+1$ & $\cdots$ & $N_S$ & $N_S+1$ 
& $\cdots$ & $N_S+N_B/2+1$ & $\cdots$ & $N_S+N_B$ \\ \hline
$\;\;\nu_i^2\;\;$ & $\tilde{\Omega}_{-N_S/2}^2$ & $\cdots$ & $\nu_{+}^2$ & $\cdots$
& $\tilde{\Omega}_{N_S/2-1}^2$ & $\tilde{\omega}_{-N_B/2}^2$ & $\cdots$ 
& $\nu_{-}^2$ & $\cdots$ & $\tilde{\omega}_{N_B/2-1}^2$ \\ \hline
\end{tabular}
\end{center}
where
\begin{eqnarray}
\nu_{\pm}^2 &=&
\frac{1}{2}\left[ \tilde{\Omega}_0^2+\tilde{\omega}_0^2
\pm \sqrt{(\tilde{\Omega}_0^2-\tilde{\omega}_0^2)^2+ \frac{4 N_S N_B c^2}{M m}}
\right],
\label{eq:Y11}
\end{eqnarray}
with
\begin{eqnarray}
M \tilde{\Omega}_s^2 &=& D + M \Omega_s^2
%+ 4 K \sin^2 \left(\frac{\pi s}{N_S} \right) 
+ c N_B, 
\label{eq:D4}\\
m \tilde{\omega}_r^2 &=& m \omega_r^2 + c N_S, 
\label{eq:D5} \\
\Omega_s &=& \sqrt{\frac{4 K}{M}} \sin \left(\frac{\pi s}{N_S} \right)
\hspace{1cm}
\mbox{for $s=-\frac{N_S}{2}, -\frac{N_S}{2}+1,\cdot\cdot\cdot, \frac{N_S}{2}-1$}, 
\label{eq:D6}\\
\omega_r &=& \sqrt{\frac{4 k}{m}} \sin \left(\frac{\pi r}{N_B} \right)
\hspace{1cm}
\mbox{for $r=-\frac{N_B}{2}, -\frac{N_B}{2}+1,\cdot\cdot\cdot, \frac{N_B}{2}-1$}.
\label{eq:D7}
\end{eqnarray}
%where $\Omega_s$ is given by Eq. (\ref{eq:B6}).
Effective frequencies of $\tilde{\Omega}_s$ and $\tilde{\omega}_r$ 
are increased with increasing $c$.
In the case of no couplings ($c=0$), eigenfrequencies are given 
\begin{center}
\begin{tabular}{|c|| c|c|c|c|c|c|c|c|c|c|} \hline
$\;\;i\;\;$ & $1$ &$\cdots$ & $N_S/2+1$ & $\cdots$ & $N_S$ & $N_S+1$ 
& $\cdots$ & $N_S+N_B/2+1$ & $\cdots$ & $N_S+N_B$ \\ \hline
$\;\;\nu_i^2\;\;$ & $\Omega_{-N_S/2}^2$ & $\cdots$ & $\Omega_{0}^2$ & $\cdots$
& $\Omega_{N_S/2-1}^2$ & $\omega_{-N_B/2}^2$ & $\cdots$ 
& $\omega_{0}^2$ & $\cdots$ & $\omega_{N_B/2-1}^2$ \\ \hline
\end{tabular}
\end{center}
The system specific heat is expressed in terms of obtained eigenfrequencies $\{ \nu_i \}$,
\begin{eqnarray}
C_S &=& k_B \sum_{i=1}^{N_S+N_B} 
\left[\frac{\beta \hbar \nu_i}
{2 \sinh (\beta \hbar \nu_i/2 )} \right]^2
- k_B \sum_{r=-N_B/2}^{N_B/2-1} \left[\frac{\beta \hbar \omega_r}
{2 \sinh (\beta \hbar \omega_r/2 )} \right]^2,
\label{eq:D8}\\
&\equiv& C-C_B,
\label{eq:D8b}
\end{eqnarray}
where $C$ and $C_B$ express the first and second terms, respectively, of Eq. (\ref{eq:D8}).
System specific heats in zero- and high-temperature limits are given by Eq. (\ref{eq:C7}).

We have calculated the system specific heat with $k=1.0$, $K=1.0$ and $D=0.0$, 
for which a unit of the energy is given by $\epsilon_0=\hbar \sqrt{k/m}$.
Figure \ref{fig7}(a) shows the temperature dependence of $C_S/k_B N_S$
for various $N_S$ with $N_B=100$ and $c_0=5.0$.
For $c_0=0.0$, $C_S$ is positive, reducing to zero at $T = 0.0$ [see Fig. 1(b)].
When $c_0$ is introduced, $C_S$ at $T \simeq 0.0$ becomes negative.
Magnitudes of negative $C_S$ are increased with decreasing $N_S$.
In order to study an origin of the negative $C_S$, we plot
$C$, $C_B$ and $C_S$ for $N_S=4$, $N_B=20$ and $c_0=5.0$ in Fig. \ref{fig7}(b),
whose inset shows eigenfrequencies for $c_0=0.0$ (open circles) and
$c_0=5.0$ (filled circles). For $c_0=0.0$, eigenfrequencies of bath and system
display dispersion relations given by Eqs. (\ref{eq:D6}) and (\ref{eq:D7}). 
By an introduced system-bath coupling of $c_0=5.0$, eigenfrequencies
are increased as shown by filled circles in the inset.
This modification in eigenfrequencies leads to a negative $C_S$ 
at $0 < k_B T/\epsilon_0 < 0.386$ where $C < C_B$. 

%\newpage
\section{Conclusion}
The specific heat has been studied of a small quantum system
consisting of $N_S$-body HOs which is strongly coupled 
to an $N_B$-body non-dissipative HO bath with identical frequency 
[Eq. (\ref{eq:B19})] or with dispersed frequencies [Eq. (\ref{eq:D7})].
The obtained results are summarized as follows:

\noindent
(i) Although the system specific heat of $C_S(T)$ is proportional to $N_S$
(extensive) in the high-temperature limit, it is not (non-extensive) at low temperatures
except for a vanishing system-bath coupling, and

\noindent
(ii) $C_S(T)$ displays an anomalous temperature dependence 
at low temperatures where it may become negative for a strong system-bath coupling.

\noindent
The item (i) implies that it is necessary to take into account finite $N_S$
for a study of small quantum open system, although previous studies have been made 
by exclusively using the CL-type models with $N_S=1$.
The item (ii) is in contrast to the result of Ingold, H\"{a}nggi and Talkner \cite{Ingold09} 
who reported that the specific heat of an HO system with $N_S=1$ 
cannot be negative although that of a free damped particle system may be negative.
They claimed that it is due to a larger specific heat of an HO by a factor of two
than that of a free particle \cite{Ingold09}.
The elucidation of non-negative specific heat of a HO system in Ref. \cite{Ingold09} 
which is valid for $N_S=1$, cannot be applied to the case of arbitrary $N_S$.

The system specific heat of $C_S=C-C_B$ given by (\ref{eq:C4b}) or (\ref{eq:D8b}) 
shows that $C_S$ expresses a change of the specific heat when the heat bath
is enlarged by coupling it to system degree of freedom.
The appearance of a negative specific heat in the item (ii) is attributed 
to increased eigenfrequencies when the system is coupled to the bath, 
by which the specific heat of $C$ is suppressed as $C < C_B$ 
and then $C_S < 0$ at low temperatures. The expression of $C_S=C-C_B$ reconciles 
with the experimental procedure to determine the specific heat 
of the system attached to the bath. First one measures the specific heat of 
the empty container and then subtracts this value from the measured specific heat 
of the combined system-plus-bath to finally obtain the specific heat of the system
\cite{Hanggi06,Hanggi08,Ingold09}. Specific heat anomalies which arise
from an attachment of the system to the environment may be common in small open 
quantum systems. It would be interesting to examine the items (i) and (ii) 
by relevant experiments for small-scale systems.

\begin{acknowledgments}
This work is partly supported by
a Grant-in-Aid for Scientific Research from 
Ministry of Education, Culture, Sports, Science and Technology of Japan.  
\end{acknowledgments}

%\vspace{0.5cm}
\appendix*

\section{A. Eigenfrequencies of $(1+1)$ and $(2+1)$ models}
\renewcommand{\theequation}{A\arabic{equation}}
\setcounter{equation}{0}

\subsection{The $(1+1)$ model}
%We may exactly solve $(1+1)$ and $(2+1)$ models.
From Eqs. (\ref{eq:A0})-(\ref{eq:A4}), the Hamiltonian for the case of 
$N_S=N_B=1$ is given by \cite{Hasegawa11a}
\begin{eqnarray}
H &=& \frac{P^2}{2M}+ \frac{D Q^2}{2}
+ \frac{p^2}{2m}+ \frac{m \omega_0^2 q^2}{2}
+\frac{c}{2}(Q-q)^2,
\label{eq:X4}
\end{eqnarray}
for which we easily obtain eigenfrequencies,
\begin{eqnarray}
\nu_{i}^2 &=& \frac{1}{2}\left[ \left( \frac{D+c}{M}+\frac{m\omega_0^2+c}{m} \right)
\pm \sqrt{\left(\frac{D+c}{M}-\frac{m\omega_0^2+c}{m} \right)^2+ \frac{4 c^2}{Mm} }
\right]
\hspace{0.3cm}\mbox{for $i=1,2$}.
\label{eq:X2}
\end{eqnarray}
The CL-type model employed in Ref. \cite{Ingold09} is given by
\begin{eqnarray}
H' &=& \frac{P^2}{2M}+ \frac{M \Omega^2 Q^2}{2}
+ \frac{p^2}{2m}+\frac{f_B}{2}(q-Q)^2,
\label{eq:X5}
\end{eqnarray}
which is related with the Hamiltonian of Eq. (\ref{eq:X4}) by $D=M \Omega^2$,
$c=f_B$ and $\omega_0=0$.
Equation (\ref{eq:X2}) is equivalent to Eq. (17) in Ref. \cite{Ingold09}.

%The CL model for $N_B=1$ is given by \cite{Caldeira81,Caldeira83}
%\begin{eqnarray}
%H_{CL} &=& \frac{P^2}{2M}+ \frac{M \Omega^2 Q^2}{2}
%+ \frac{p^2}{2m}+\frac{m \omega_0^2}{2}\left(q-\frac{c Q}{m \omega_0^2} \right)^2,
%\end{eqnarray}
%for which we obtain eigenfrequncies,
%\begin{eqnarray}
%\nu_{i}^2 &=& \frac{1}{2}\left[ \left(\Omega^2+\omega_0^2+\frac{2 c^2}{Mm\omega_0^2}\right)
%\pm \sqrt{\left(\Omega^2-\omega_0^2 \right)^2
%+ \frac{2 c(\Omega^2+\omega_0^2)}{Mm \omega_0^2} }
%\right]
%\hspace{0.3cm}\mbox{for $i=1,2$}.
%\end{eqnarray}

\subsection{The $(2+1)$ model}
The Hamiltonian for the case of $N_S=2$ and $N_B=1$ is given by %Eq. (\ref{eq:A0}) 
$H=H_S+H_B+H_I$ with
\begin{eqnarray}
H_S &=& \frac{1}{2M}(P_1^2+P_2^2)+\frac{D}{2}(Q_1^2+Q_2^2)
+\frac{J}{2}(Q_1-Q_2)^2, \\
H_B &=& \frac{p^2}{2m}+\frac{m \omega_0^2 q^2}{2}, \\
H_I&=& \frac{c}{2}\left[(q-Q_1)^2+(q-Q_2)^2 \right],
\end{eqnarray}
meanings of all terms being trivial.
Introducing center-of-mass and relative variables,
\begin{eqnarray}
Q_{c} &=& \frac{1}{2}(Q_1+Q_2),\;\;Q_r =Q_1-Q_2, \\
P_c &=& P_1+P_2, \;\;\; P_r = \frac{1}{2}(P_1-P_2),
\end{eqnarray}
we obtain $H_S$ and $H_I$ given by
\begin{eqnarray}
H_S &=& \frac{P_c^2}{2 M_c}+ D Q_c^2
+\frac{P_r^2}{2 M_r}+\frac{D Q_r^2}{4}+\frac{J Q_r^2}{2}, \\
H_I &=& c\left( q^2-2 q Q_c+Q_c^2+\frac{Q_r^2}{4} \right),
\end{eqnarray}
where $M_c=2 M$ and $M_r=M/2$.
Equations of motion for $Q_c$, $Q_r$ and $q$ are given by
\begin{eqnarray}
M_c \ddot{Q}_c &=& -2(D+c) Q_c + 2c q, \\
M_r \ddot{Q}_r &=& - \frac{1}{2}(D+2J+c) Q_r, \\
m \ddot{q} &=& -(m \omega_0^2+2 c)q+2 c Q_c.
\end{eqnarray}
We note that $Q_c$ is coupled with $q$ whereas $Q_r$ is decoupled
from $Q_c$ and $q$.
A simple calculation leads to three eigenfrequencies given by
\begin{eqnarray}
\nu_i^2 &=& \left\{ \begin{array}{ll}
\frac{1}{M}(D+2 J+c)
\quad & \mbox{for $i=1$}, \\ 
\frac{1}{2}\left[\left( \frac{D+c}{M}+\frac{m \omega_0^2 + 2 c}{m} \right)
\pm \sqrt{\left( \frac{D+c}{M}-\frac{m \omega_0^2 + 2 c}{m} \right)
+ \frac{8 c^2}{M m}}
\right]
\quad & \mbox{for $i=2, 3$}. 
\label{eq:X1}
\end{array} \right. 
\end{eqnarray}
Equation (\ref{eq:X1}) agrees with Eq. (\ref{eq:B18}) for $N_S=2$ and $N_B=1$
because $M \tilde{\Omega}_0^2=D+c$, $m \tilde{\omega}_0^2=m \omega_0^2+2 c$ 
and $J=2 K$ which is due to a double counting of interactions for $N_S=2$:
$K\sum_{k=1}^2 (Q_k-Q_{k+1})^2=K[(Q_1-Q_2)^2+(Q_2-Q_1)^2]
=J(Q_1-Q_2)^2$.

It is evident from Eqs. (\ref{eq:C4}), (\ref{eq:X2}) and (\ref{eq:X1}) that we obtain
\begin{eqnarray}
C_S(T; 2,1) &\neq& 2\:C_S(T; 1,1)
\hspace{2cm}\mbox{for $T \simeq 0.0$},
\label{eq:X3} \\
&=&  2\:C_S(T; 1, 1)=2 k_B
\hspace{1cm}\mbox{for $T \rightarrow \infty$},
\end{eqnarray}
where $C_S(T; 1,1)$ and $C_S(T; 2,1)$ denote $T$-dependent system specific heats for
the $(1+1)$ model and $(2+1)$ model (with $J=0.0$), respectively.

\section{B. Baths consisting of coupled harmonic oscillators}
\renewcommand{\theequation}{B\arabic{equation}}
\setcounter{equation}{0}
We consider an $N_S$-body coupled system subjected to an $N_B$-body coupled bath,
whose Hamiltonian is given by Eqs. (\ref{eq:A0}), (\ref{eq:A2}), (\ref{eq:A4}) 
and (\ref{eq:D1}).
Applying the canonical transformation given by Eqs. (\ref{eq:B2}) and (\ref{eq:B3}) 
to the system with the boundary condition given by Eq. (\ref{eq:B1}), we obtain
$H_S$ given by Eqs. (\ref{eq:B4}) and (\ref{eq:B6}).
When we apply the canonical transformation \cite{Hasegawa11a,Florencio85},
\begin{eqnarray}
q_n &=& \frac{1}{\sqrt{N_B}} \sum_{r=-N_B/2}^{N_B/2-1}
e^{i (2 \pi n r/N_B)} \tilde{q}_r, 
\label{eq:Y1}\\
p_n &=& \frac{1}{\sqrt{N_B}} \sum_{r=-N_B/2}^{N_B/2-1}
e^{i (2 \pi n r/N_B)} \tilde{p}_r,
\label{eq:Y2}
\end{eqnarray}
to the bath with the periodic condition given by Eq. (\ref{eq:D2}),
$H_B$ in Eq. (\ref{eq:D1}) becomes
\begin{eqnarray}
H_B &=& \sum_{r=-N_B/2}^{N_B/2-1} \left(\frac{\tilde{p}_r^* \tilde{p}_r}{2m}
+ \frac{m \omega_r^2 \tilde{q}_r^* \tilde{q}_r}{2} \right), 
\label{eq:Y3}
\end{eqnarray}
with
\begin{eqnarray}
\omega_r^2 &=& \left( \frac{4k}{m}\right) \sin^2 \left(\frac{\pi r}{N_B} \right)
\hspace{1cm}
\mbox{for $r=-\frac{N_B}{2}, -\frac{N_B}{2}+1,\cdot\cdot\cdot, \frac{N_B}{2}-1$},
\label{eq:Y4}
\end{eqnarray}
where the commutation relations: 
\begin{eqnarray}
[\tilde{q}_r, \tilde{p}_{r'}^*]=i \hbar \delta_{r r'}, \;\;\;
[\tilde{q}_r, \tilde{q}_{r'}^*]=[\tilde{p}_r, \tilde{p}_{r'}^*]=0,
\end{eqnarray}
hold with $\tilde{q}_r^*=\tilde{q}_{-r}$ and $\tilde{p}_r^*=\tilde{p}_{-r}$.

By the canonical transformations given by Eqs. (\ref{eq:B2}), (\ref{eq:B3}),
(\ref{eq:Y1}) and (\ref{eq:Y2}), $H_I$ in Eq. (\ref{eq:A4}) becomes
\begin{eqnarray}
H_I &=& \frac{c N_B}{2} \sum_{s=-N_S/2}^{N_S/2-1} \tilde{Q}_s^*\tilde{Q}_s
+\frac{c N_S}{2} \sum_{r=-N_B/2}^{N_B/2-1} \tilde{q}_r^2
-c \sqrt{N_S N_B}\:\tilde{Q}_0 \:\tilde{q}_0.
\label{eq:Y5}
\end{eqnarray}

Equations of motion for $\tilde{Q}_k$ and $\tilde{q}_r$ are expressed by
\begin{eqnarray}
M \ddot{\tilde{Q}}_s &=& - M \tilde{\Omega}_s^2 \: \tilde{Q}_s 
\hspace{4cm} \mbox{for $s \neq 0$}, 
\label{eq:Y7}\\
M \ddot{\tilde{Q}}_0 &=& - M \tilde{\Omega}_0^2 \: \tilde{Q}_0
+ c \sqrt{N_S N_B} \:\tilde{q}_0
\hspace{1cm} \mbox{for $s=0$}, 
\label{eq:Y8}\\
m \ddot{\tilde{q}}_r &=& - m \tilde{\omega}_r^2 \:\tilde{q}_r
\hspace{4cm} \mbox{for $r \neq 0$}, 
\label{eq:Y9}\\
m \ddot{\tilde{q}}_0 &=& - m \tilde{\omega}_0^2 \:\tilde{q}_0 
+ c \sqrt{N_S N_B} \:\tilde{Q}_0
\hspace{1cm} \mbox{for $r = 0$},
\label{eq:Y10}
\end{eqnarray}
with
\begin{eqnarray}
M \tilde{\Omega}_s^2 &=& D + 4 K \sin^2 \left(\frac{\pi s}{N_S} \right) + N_B \:c, \\
m \tilde{\omega}_r^2 &=& 4 k \sin^2 \left(\frac{\pi r}{N_B} \right) + N_S \:c.
\end{eqnarray}
Equations (\ref{eq:Y7})-(\ref{eq:Y10}) show that $\tilde{Q}_0$ and $\tilde{q}_0$ 
are coupled although they are decoupled from the rest of variables. 
From Eqs. (\ref{eq:Y7})-(\ref{eq:Y10}), we obtain eigenfrequencies 
which have been presented in Sec. III B. 

%\newpage

\newpage
\begin{figure}
\begin{center}
\end{center}
\caption{
%(Color online) 
Temperature dependences of energy and specific heat of a coupled HO system and 
an uncoupled HO bath with $c_0=0.0$.
(a) $E_B/N_B$ (dashed curve), and $E_S/N_S$ for various $N_S$:
$N_S=4$ (chain curves), $10$ (dotted curves), $100$ (double-chain curve) and 1000 (solid curves).
(b) $C_B/k_B N_B$ (dashed curve), and $C_S/k_B N_S$ for various $N_S$ which are
same as in (a), the inset showing $C_S/k_B T N_S$ ($\epsilon_0=\hbar \omega_0$).
}
\label{fig1}
\end{figure}

\begin{figure}
\begin{center}
\end{center}
\caption{
%(Color online) 
Temperature dependences of $C_S/k_B N_S$ of coupled HO systems
subjected to uncoupled HO baths with $N_S=4$ and $N_B=100$
for various $c_0$: $c_0=0.0$ (solid curve), 
$0.01$ (dotted curve), $0.1$ (dashed curve), 0.5 (bold solid curve), 
1.0 (chain curve), 2.0 (bold dashed curve)
and 5.0 (double-chain curve).
}
\label{fig2}
\end{figure}

\begin{figure}
\begin{center}
\end{center}
\caption{
%(Color online) 
Temperature dependences of $C_S/k_B N_S$ of coupled HO systems
subjected to uncoupled HO baths for (a) $c_0=0.1$, (b) $c_0=1.0$
and (c) $c_0=5.0$ with $N_B=100$ for various $N_S$: 
$N_S=4$ (solid curve), $10$ (dashed curve) and 100 (chain curve).
}
\label{fig3}
\end{figure}

\begin{figure}
\begin{center}
\end{center}
\caption{
%(Color online) 
Temperature dependences of $C_S/k_B N_S$ of coupled HO systems
subjected to uncoupled HO baths with $N_S=4$ for various $N_B$: 
$N_B=4$ (solid curve), $10$ (dotted curve), $100$ (dashed curve) and 1000 (chain curve) 
with $c_0=0.1$, $1.0$ and $5.0$. 
Results for $N_B=4$, 10, 100 and 1000 with $c_0=0.1$ are indistinguishable.
}
\label{fig4}
\end{figure}

\begin{figure}
\begin{center}
\end{center}
\caption{
%(Color online) 
Temperature dependences of $C/k_B $ (chain curve), $C_B/k_B $ (dashed curve)
and $C_S/k_B $ (solid curve) of coupled HO systems
subjected to uncoupled HO baths for $c_0=5.0$, $N_S=4$ and $N_B=10$:
$C$ and $C_B$ denote specific heats of system-plus-bath and bath, respectively,
and the system specific heat is given by $C_S=C-C_B$. 
The inset shows eigenfrequencies $\nu_i$ for $c_0=0.0$ (open circles)
and $c_0=5.0$ (filled circles), dashed lines being plotted for a guide of the eye 
(see text).
}
\label{fig5}
\end{figure}

\begin{figure}
\begin{center}
\end{center}
\caption{
%(Color online) 
Temperature dependences of specific heat of uncoupled HO systems
subjected to uncoupled HO baths with $\omega_0=1.0$, $D=1.0$ and $K=0.0$.
(a) $C_S/k_B N_S$ with $c_0=5.0$ and $N_B=100$ for various $N_S$: 
$N_S=1$ (solid curve), $2$ (dashed curve), $4$ (dotted curve) and 10 (chain curve).
(b) $C/k_B $ (chain curve), $C_B/k_B $ (dashed curve) and $C_S/k_B $ (solid curve)
for $c_0=5.0$, $N_S=4$ and $N_B=10$:
the inset shows eigenfrequencies $\nu_i$ for $c_0=0.0$ (open circles)
and $c_0=5.0$ (filled circles), dashed lines being plotted for a guide of the eye 
($\epsilon_0=\hbar \omega_0$).
}
\label{fig6}
\end{figure}

\begin{figure}
\begin{center}
\end{center}
\caption{
%(Color online)
Temperature dependences of specific heats of coupled HO systems subjected
to coupled HO baths expressed by Eq. (\ref{eq:D1}) with $k=1.0$, $K=1.0$ and $D=0.0$.
(a) $C_S/k_B N_S$ with $c_0=5.0$ and $N_B=100$ for various $N_S$:
$N_S=4$ (solid curve), $10$ (dashed curve) and 100 (chain curve). 
(b) $C/k_B $ (chain curve), $C_B/k_B $ (dashed curve) and $C_S/k_B $ (solid curve)
for $c_0=5.0$, $N_S=4$ and $N_B=20$:
the inset shows eigenfrequencies $\nu_i$ for $c_0=0.0$ (open circles)
and $c_0=5.0$ (filled circles), dashed lines being plotted for a guide of the eye 
($\epsilon_0=\hbar \sqrt{k/m}$).
}
\label{fig7}
\end{figure}
\end{document}